\documentclass[12pt]{article} 
\usepackage[latin1]{inputenc}
\usepackage [dvips]{graphicx}
\newcommand{\ee}{\end{equation}}
\newcommand{\be}{\begin{equation}}
\begin{document}

\title {Zipf's law in Multifragmentation}
\author{X. Campi\thanks{Corresponding author: campi@ipno.in2p3.fr}  and H.
Krivine \\ \small Laboratoire de Physique Th\'eorique et Mod\`eles
Statistiques \thanks{Unit\'e de Recherche de l'Universit\'e de Paris XI
associ\'ee au CNRS (UMR 8626)}\\ \small B\^at. 100, Universit\'e de Paris
XI, F-91405 Orsay Cedex, France}

\maketitle
\begin{abstract}
 We discuss the meaning of Zipf's law in nuclear multifragmentation. We
remark that Zipf's law is a consequence of a power law fragment size
distribution with exponent $\tau \simeq 2$. We also recall why the presence
of such distribution is not a reliable signal of a liquid-gas phase
transition.

PACS number(s): 25.70.Pq, 05.70.Jk, 64.60.Ak
\end{abstract}
%\twocolumn

The search for reliable signatures of the liquid-gas phase transition in
nuclear multifragmentation is, both theoretically and experimentally, one of
the major issues of this field of physics.  The empirical observation that
the size distribution of heavier clusters generated in various processes
satisfies the so called Zipf's law \cite{Zipf}, has raised interest and
curiosity. This was first pointed out by Y.G. Ma \cite{Ma} in the framework
of the isospin dependent lattice gas model and more recently seen in nuclear
fragmentation data \cite{Ma_Exp,Dabrowska}.

In the present context, Zipf's law \footnote{In its original formulation,
Zipf's law concerns the rank of the frequency  of words in a
text.} states that the mean size (mass or charge) $\bar s(r) $ of the
largest, second-largest...$r$-largest clusters, decreases according to their
rank $r=1, 2,\cdots, n$ as \be \bar s(r) \sim 1/r^{\lambda}
\label{e:sig},\ee with $\lambda \simeq 1$.

The examination of the above mentioned numerical simulations \cite{Ma} and
experimental data \cite{Ma_Exp,Dabrowska} , shows that a fairly good
agreement with Eq. (1) is indeed obtained when the exponent is $\lambda
\simeq 1$. This happens when other observables reach extreme values (maximum
value of the moments of the cluster size distribution ($\emph{c.s.d.}$),
minimum of the effective $\tau$ parameter fit of the $\emph{c.s.d.}$,
maximum fluctuation of the largest fragment...). This seems to be the origin
of the suggestion \cite{Ma} that the fulfillment of the Zipf's law is a good
signal of the liquid gas phase transition.

The aim of this note is to point out that the finding of a Zipf's law is
nothing but a consequence of the power law shape of the $\emph{c.s.d.}$ with
exponent $\tau \simeq 2$. More precisely, both exponents are connected through
the formula $\lambda=1/(\tau-1$).

The proof of this statement is straightforward. Let be $s$ the size (mass or
charge) of the clusters and $s(r)$ the size of the cluster of rank $r$. If
the $\emph{c.s.d.}$ is a power low \be Pr[s \in (s, s+ds)] \sim ds/s^
\tau,\ee integrating over $s$, one gets the probability to find a cluster of
size larger than $S$
\be Pr[s > S] \sim 1/S^{ (\tau-1)}. \label{3}
\ee

We now take $S = \bar s(r)$, where $ \bar s(r)$ is the average size of
clusters of rank $r$. This is a strictly decreasing real function of $r$,
hence without degeneracy. On an infinite sampling, the event $E$, ``one
cluster randomly chosen has a size larger than $ \bar s(r)$'' is identical
to the event ``his rank is larger than $r$''. Arranging in ascending
order the ranks from $1$ to $n$, the probability of $E$ is 
\be Pr(E) = Pr[s > \bar s(r)] = \frac{r-1}{n-1} \sim r. \ee On the other hand,
Eq.(\ref{3}) gives now
$$ Pr(E)\sim 1/\bar s(r)^{\tau-1}.$$
Therefore
$$ r\sim 1/\bar s(r)^{\tau-1},$$
and from Eq. (1),
\be \lambda = 1/(\tau - 1) \label{4}.\ee

Similar proofs can be found in the literature (see for example
ref. \cite{Proof}, where the same arguments are used to proof that if
the ranking follows Eq. (1), then the \emph{c.s.d.} is necessarily a power law).

  The above formulae are strictly valid for infinite samplings. We have
checked numerically that these remain accurate for finite samplings. We
proceeded as follows. We generate partitions of an integer number $N$, with
the condition that the mean $\emph{c.s.d.}$ is a power law of given exponent
$\tau$ (i.e. each part of $N$ is taken as a cluster size $s$). From these
random numbers $s$, we construct the function $\bar s(r)$. As expected, for
large $N$ ($N > 1000$), the function $ \bar s(r)$ is very close to a perfect
power law in a large domain of $r$ and Eq.(\ref{4}) is satisfied within
numerical uncertainties in the fits of the power laws. For $N \simeq 100$
and $2\le \tau \le 3$, Eq.(\ref{4}) is fulfilled within a few percent. In
general, $\bar s(1)$ lies above the best fit curve. This is due to the
finite size. Indeed, in the domain of $s$ contributing to $\bar s(1)$, the
\emph{c.s.d.} deviates from a power law. For the same reasons, for larger
values of $\tau$ the agreement is less good.  Fig.1 shows an example of
$\bar s(r)$ calculated from a power law \emph{c.s.d.} with $\tau=2.2$. A
best fit with Eq. (1) gives $\lambda \simeq 0.86$.
 
\begin{figure}[h!] 
\begin{center}
\includegraphics[scale=0.4,angle=-90]{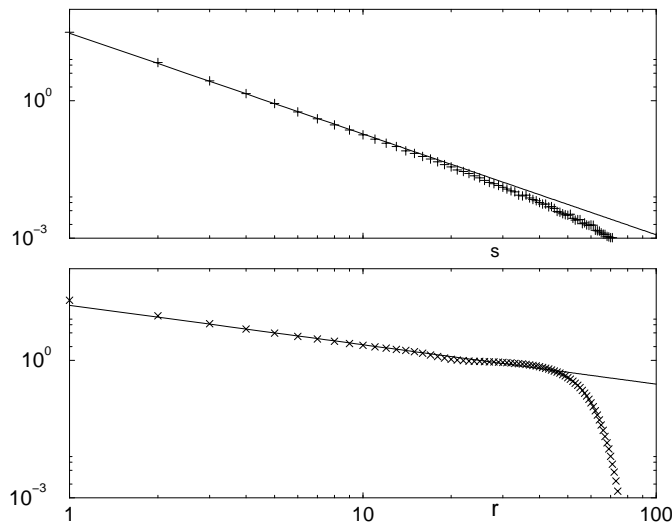}
\end{center}
\caption{\label{s1} \it Cluster size distribution $n(s)$ (upper panel) and
mean size $\bar s(r)$ distribution of clusters of rank $r$ (lower panel),
corresponding to a power law $\emph{c.s.d.}$ with $\tau=2.2$. The size $s$
of the clusters is generated from random partitions of N=100 with the
constraint that the mean \emph{c.s.d.} is a power law with exponent
$\tau=2.2$ (see text). The straight lines are best fits, with slopes
$\tau=2.2$ and $\lambda=0.86$, respectively.}
\end{figure}
\begin{figure}[h!] 
\begin{center}
\includegraphics[scale=0.4,angle=-90]{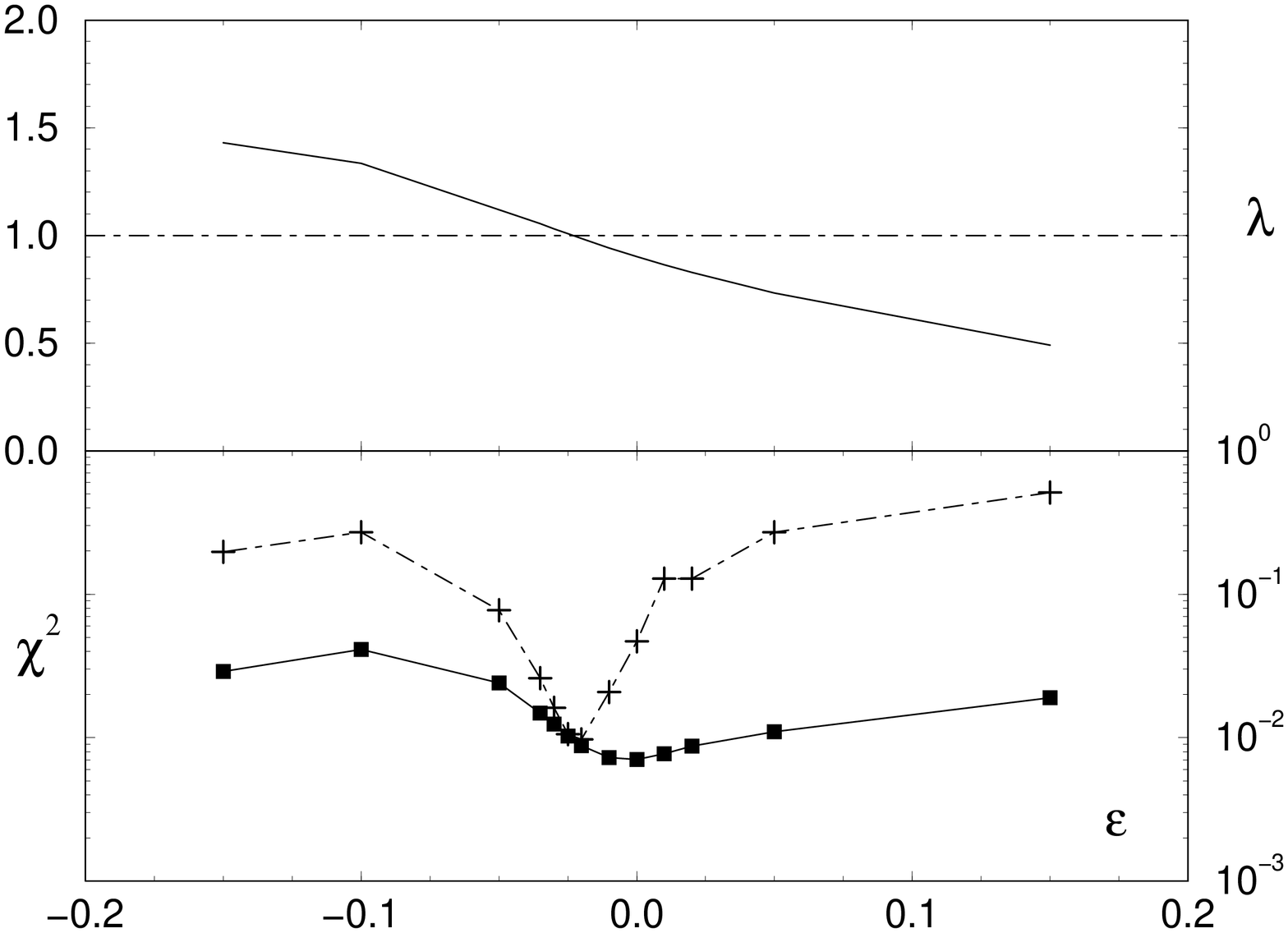}
\end{center}
\caption{\label{s2} \it The slope parameter $\lambda$ (Eq. (1))(upper
panel) and the $\chi^2$ deviation from a power law of $\bar s(r)$ (lower
panel), as a function of the distance $\epsilon $ to the percolation
critical point (see text). The continuous line (squares) corresponds to the
best fit with $\lambda $ free and the dashed-dot line (crosses) with $\lambda=1$ 
\cite{Ma}. }
\end{figure}

It is interesting to see what happens if the $\emph{c.s.d.}$ is not a power law.
For example, for an exponential distribution $ \sim \exp (- \alpha s)$,
following the same reasoning, one finds $ \bar s (r) \sim -\log r$. Various
theories of cluster formation \cite{Lattice-gas,Stauffer} offer
other interesting examples. Close to their  critical point, the
$\emph{c.s.d.}$ behaves as
 \be n(s) \sim s^{- \tau} f(s^{\sigma} \epsilon ), \label{6}\ee
where $f$ is a scaling function, $\epsilon$ the distance to the critical
point and $\tau, \sigma$ two critical exponents. Right at the critical
point, the scaling function $f(0)=1$.  In three dimensions $\tau \simeq
2.21$ for the lattice-gas model with clusters defined according to the
Coniglio-Klein prescription \cite{Lattice-gas,C-K} and $\tau \simeq 2.18$ for
percolation theory \cite{Stauffer}. As expected, (approximate) Zipf's laws
have been observed \cite{Ma,Batanabe} in the vicinity of the corresponding
critical points. 

We present below some results for a bond-percolation calculation
\cite{Stauffer} with $N=100$ occupied sites. In Fig. (2) we show as a
function of the distance to the critical bond parameter $\epsilon=p_{c}-p$,
the evolution of $\lambda$ (upper panel) and the $\chi²$ deviation of the
$\bar s(r)$ with respect to two power law fits (lower panel). The continuous
line (squares) is the best fit of $\bar s(r)$ with two parameters
$c/r^{\lambda}$. The minimum of $\chi²$ occurs when $\lambda \simeq 0.90$ for a
slope $\tau \simeq 2.16$ of the corresponding \emph{c.s.d.}, in agreement
with Eq.(\ref{4}) within $4 \%$.  The dashed-dotted line (crosses)
corresponds to the fit with fixed $\lambda=1$ (i.e. a strict Zipf's law, as
done in ref. \cite{Ma}). This fit, that violates Eq.(\ref{4}) by $26 \% $,
giving an incorrect localization of the critical point, shows as expected a
larger $\chi^2$.  Similarly, for the lattice-gas model with Coniglio-Klein
clusters one expects, according to Eq.(1), $ \lambda \simeq 0.83$ \cite{Ma}.

Before to summarize, we would like to add a comment on the observation of
power law $\emph{c.s.d.}$ and Zipf's laws. As mentioned before, various
theories of cluster formation (Fischer droplets \cite{Fisher}, lattice-gas
\cite{Lattice-gas}, Lennard-Jones fluids with Hill's clusters \cite{C.K.S.},
percolation \cite{Stauffer}), predict power laws $\emph{c.s.d.}$ with $\tau
\simeq 2$ at the corresponding critical points. However, such
$\emph{c.s.d.}$ also appear elsewhere (for example, in the super-critical
region of the lattice-gas and realistic Lennard-Jones fluids).

In summary, (approximate) Zipf's law is just a mathematical consequence of a
 power law \emph{c.s.d.} with exponent $\tau \simeq 2$. Such distributions appear
 at the critical point of many theories, but also elsewhere. In consequence,
 we conclude that the observation of a Zipf's law is neither a new and
 independent signal of a critical behavior, nor an unambiguous signal of a
 thermodynamical phase transition.

We thank Marc M\'ezard for a very helpful discussion.

\end{document}